# Investigating Concerns of Security and Privacy Among Rohingya Refugees in Malaysia

Theodoros Georgiou*

Heriot-Watt University, t.georgiou@hw.ac.uk

Lynne Baillie

Heriot-Watt University, l.baillie@hw.ac.uk

Ryan Shah

Heriot-Watt University, r.shah@hw.ac.uk

The security and privacy of refugee communities have emerged as pressing concerns in the context of increasing global migration. The Rohingya refugees are a stateless Muslim minority group in Myanmar who were forced to flee their homes after conflict broke out, with many fleeing to neighbouring countries and ending up in refugee camps, such as in Bangladesh. However, others migrated to Malaysia and those who arrive there live within the community as urban refugees. However, the Rohingya in Malaysia are not legally recognized and have limited and restricted access to public resources such as healthcare and education. This means they face security and privacy challenges, different to other refugee groups, which are often compounded by this lack of recognition, social isolation and lack of access to vital resources. This paper discusses the implications of security and privacy of the Rohingya refugees, focusing on available and accessible technological assistance, uncovering the heightened need for a human-centered approach to design and implementation of solutions that factor in these requirements. Overall, the discussions and findings presented in this paper on the security and privacy of the Rohingya provides a valuable resource for researchers, practitioners and policymakers in the wider HCI community.

CCS CONCEPTS • Security and privacy ~ Human and societal aspects of security and privacy • Human-centered computing

## 1 INTRODUCTION

The Rohingya are a stateless Muslim minority group in Myanmar, who were forced to flee their homes when conflict broke out in Myanmar's Rakhine State. Myanmar continues to deny the Rohingya citizenship and refuse to recognize them as people [1], even after the International Court of Justice ordered Myanmar to take steps to protect members of the Rohingya community from genocide. The majority of Rohingya refugees who flee Myanmar end up in refugee camps in Bangladesh

(880,000 in 2017 [4]), with others migrating to Thailand and Malaysia. Rohingya refugees who arrive in Malaysia (currently around 105,000 [2]) live within the community as urban refugees [3] – a somewhat unique situation for this stateless refugee community – with the majority concentrated around the capital and other major areas of the country, such as Penang. However, being stateless in an urban environment, as opposed to refugee camps, presents the Rohingya with its own set of challenges. Specifically, they are in a Catch-22 situation, where they cannot go back to where they are from (i.e. Myanmar) as they are not given papers to prove they are genuine refugees of Myanmar, nor can they move forward or get statehood where they are, again due to the lack of said papers. In Malaysia, for example, there is no ratification of the 1951 Refugee Convention or the 1967 Protocol [5], meaning the Rohingya are not legally recognized and have limited/restricted access to public resources such as public healthcare and education [6] and having no paperwork proving their status hinders their legal recognition further. This demonstrates the uniqueness of the stateless position of Rohingya refugees, in comparison with other refugee groups.

Having fled their own country to avoid extreme violence, it is no surprise that one of main concerns pertaining to the Rohingya community surround safety and functions that can be used for an emergency. Unfortunately, Rohingya people are subject to abuse, violence and exploitation in Malaysia, with authorities such as the police often facilitating this kind of behaviour. For example, the police may arrest Rohingya people and are they told to hand over whatever they have. Given their status of being stateless, they would likely feel they have no other option but to hand over their money or mobile phone. Often living hand to mouth and working illegally, the money they do have is a crucial lifeline that supports them and thus privacy concerns of tracking and profiling are likely most important to mitigate and avoid exploitation. Therefore, it is vital to assume that existing pathways designed for the safety of the general Malay population, may not be appropriate for Rohingya people and alternative means of emergency contacts, for example, are required. Furthermore, there is a lack of readily available and accessible technological assistance that can aid urban refugee populations like the Rohingya [7]. While many applications are available for communicating a need for safety or aid, these often do not correlate with the needs of illegal, undocumented immigrants in marginalised communities [8,9], and in particular to stateless, marginalised groups in urban environments like the Rohingya [10-13]. For example, families or subsets of the urban Rohingya community often have to share a mobile phone, resulting in concerns surrounding personal privacy and consent left unanswered. Existing consent protocols for sharing information, such as privacy policies for social media applications like TikTok, are designed with individual or personal usage of the platform in mind and will likely not be suitable for the stateless and marginalised.

## 2 METHODOLOGY

We carried out a multiphase workshop (discussed in detail in [14]) that aimed to gain a broader understanding of the needs of the Rohingya refugees from technologies they already have access to (e.g. mobile phones) and how new technologies utilising these technologoies can provide assistance to daily activities, both early on and long after they arrive in a new host country, while understanding the experiences and challenges faced by them [15,16,17]. In this paper, we take a deep look into the security and privacy concerns of refugees living in an urban environment. This is facilitated by data and insights we gathered through the multiphase workshop we carried out with Rohingya refugees in Malaysia and the NGO volunteers who support them. In total, 54 Rohingya (27 male and 27 female) ranging from 13 to 45 years old participated in the study. Due to relatively high illiteracy levels within the Rohingya community, only verbal consent was taken from all participants to avoid making people feel excluded because they may not be able to read or write and we did not explicitly ask them to write anything. The NGO handled recruitment and consent and made it explicitly clear that participation would not affect their relationship or the services and support they provide to any individual or the community. The procedure for this study and all ethics proposals were written in collaboration with a local NGO (whose mission is ensuring the safety of the Rohingya community) and our Malaysian expert in ethnic minority and refugee communities collaborator. Several of the NGO staff are also Rohingya themselves, hence they fully understand the needs of their community. Ethics was granted by the NGO's ethics review board (Malaysia) and our university's ethics committee (UK). Minors were either accompanied by a parent or the NGO when parent accompaniment was not possible. The initial phases of the study aimed at acquiring information about: experiences pertaining to living situations when they first arrived as refugees in Malaysia to the current day; what technologies they had access to when first arriving as refugees and how they use it in the present day; and finally, to design and discuss new technology concepts (e.g. an app or interaction) based on their experiences as stateless refugees. Then participants of that study were given an activity with cultural probes [18,19] in order to gain insights into the unique and personal circumstances of participants to facilitate future designs.

## 2.1 Phase 1 – Participatory Design Workshop

Within the first phase, discussions were focused on the current situation and the integration and influence technology has on their every-day life. Being stateless, Rohingya refugees, means they are subject to harassment, abuse, violence and exploitation – with authorities such as the police often facilitating this kind of behaviour. Most of the Rohingya live hand to mouth, working illegally in order to support themselves and their families as the Malaysian government does not grant them working permits. Mobile phones and the use of the Internet are critical for informing family members that they are safe when they have to travel outside the community, or when confronted with dangerous situations such as exploitation by the police. For example, one participant mentioned that they "*Get to call mother when emergency. If there is someone bothering me or anything, I could just straight away call my father.*" Many of the Rohingya often share mobile phones between family or members of the community. Current consent protocols do not assist with this and may result in making them more discoverable, and easier to attack or exploit. We found that participants were very conscious of their online data sharing, with some participants drawing or writing about functions to help them better manage how and with whom they share data on social media (Figure 1). This example highlights the previous point we make about existing privacy policies and consent protocols for platforms such as TikTok understood to not be appropriate by the Rohingya.

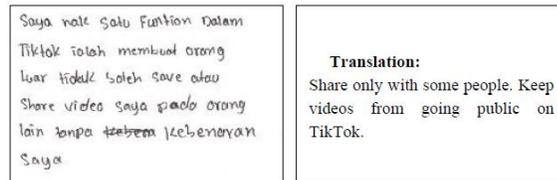

Figure 1: Privacy awareness related to existing consent protocols in social media platforms

## 2.2 Phase 2 – Cultural Probes

Each participant received a cultural probe kit, containing a sketchbook with tasks, a voice recorder and a camera, with the aim of allowing us to receive deeper insights on their background. Many of the participants viewed money as their most private object, which is clear given that the Rohingya are subject to exploitation by authorities such as the police. This is highlighted by one participant who said "*If my husband or child just go around and play around the flat, the village [...] the police come and arrest and they say give what you have [...] If you don't give, you will go to the court*." Interestingly, observing the photographs they had submitted demonstrated high-levels of knowledge pertaining privacy awareness and safety to avoid profiling and tracking which may lead to exploitation. While we never instructed participants to do so, many took pictures of their faces hidden by a number card we gave them to annotate the pictures, or by turning their backs away from the camera (Figure 2) in their own home where you would assume they feel reasonably secure. The sense of not being secure or settled was evident from these photographs, with houses they live in appearing to have very little to no personalization, while personal belongings look almost packed and "ready to go".

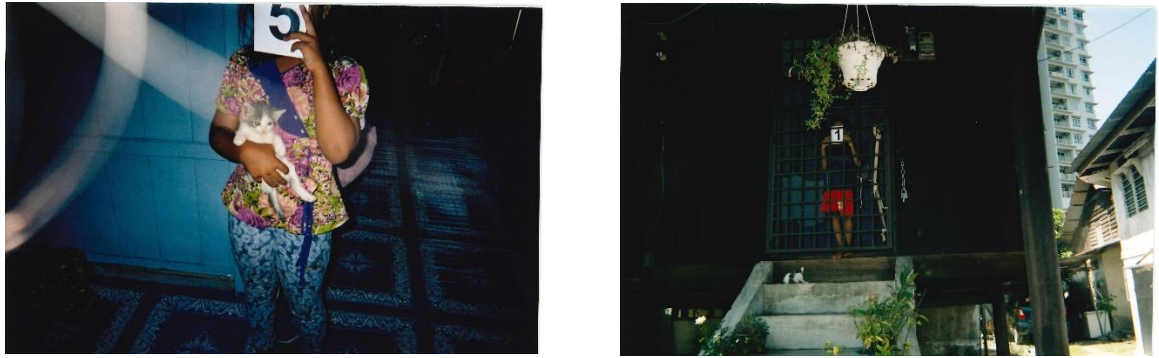

Figure 2: Participants electing to hide their face during a cultural probe activity.

## 3 DISCUSSION AND IMPACT OF SECURITY AND PRIVACY OF ROHINGYA REFUGEES

Privacy and safety are key requirements for all individuals, regardless of ethnicity or background. Marginalised urban refugees may have limited access to digital technologies and are at risk of digital exclusion that can have serious implications for their security and privacy [20]. With the Rohingya refugees being stateless and living in an urban population, the privacy and safety requirements for technologies they use, for example to contact appropriate aid organisations, are more delicate – which is not always the case for other refugees – especially for those living in organized camps. This leaves them vulnerable to harassment and exploitation by authorities such as the police, resulting in them often hiding their identity to avoid tracking and profiling (Figure 2), and keeping money or valuables secret. Other marginalised urban refugees also face exploitation by employers and landlords, with a lack of access to legal assistance and protection [21]. Further while many refugees are often not allowed to legally work, given the Rohingya are stateless and vulnerable to exploitation from authorities, they typically live in situations where they are able to quickly pack up and leave. While apps and digital tools for safety do exist, such as those which offer live geolocation of emergencies and aid facilities[1], these may not be appropriate for Rohingya people given the risk of exploitation. One solution may be to use anonymous geolocation tags to map potential hotspots to help others avoid potential exploitation and alert NGOs to initiate preventative investigations. In any case, special consideration must be given to tools that make use of personal identifiers (e.g. geolocation, names, etc.). This must be done to mitigate for tracking and profiling, while existing consent protocols must take unique circumstances into account and in a manner the user understands and has consented to. This is highlighted by the high levels of privacy awareness demonstrated by the Rohingya participants. Ultimately, our findings can benefit other ethnic minority communities globally, as well as the Rohingya. The wider HCI community can learn from the presented discussions on security and privacy for unique minorities and marginalised communities, building on findings and experiences while drawing conclusions applicable to communities they investigate.

---

[1] https://www.safezoneapp.com